\documentclass[11pt]{article}
\usepackage{amssymb,amsmath,epsfig}
\makeatletter

\@addtoreset{equation}{section}
\def\section{\@startsection {section}{1}{\z@}{-2.5ex plus -1ex minus
 -.2ex}{1.3ex plus .2ex}{\large\bf}}
\def\subsection{\@startsection{subsection}{2}{\z@}{-2.25ex plus%
 -1ex minus -.2ex}{0.5ex plus .2ex}{\bf}}

\advance \voffset by -0.9in
\advance \hoffset by -0.6in
\newcommand{\inv}[0]{{-1}}

\newcommand{\cif}[0]{\mathcal{C}^\infty}

\newcommand{\hyp}[0]{{\mathbb{H}^2}}

\def\ba{{\mbox{\boldmath $a$}}}

\def\bx{{\mbox{\boldmath $x$}}}

\def\bp{{\mbox{\boldmath $p$}}}

\def\bq{{\mbox{\boldmath $q$}}}

\newcommand{\RR}{\mathbb{R}}

\newtheorem{theorem}{Theorem}[section]

\def\bea{\begin{eqnarray}}
\def\eea{\end{eqnarray}}
\def\bmz{\left(\begin{array}{2,2}}
\def\emz{\end{array}\right)}
\def\bmd{\left(\begin{array}{3,3}}
\def\emd{\end{array}\right)}

\def\bpm{\begin{pmatrix}}
\def\epm{\end{pmatrix}}

\begin{document}
\parskip 6pt
\parindent 0pt

\begin{center}
\baselineskip 24 pt
{\Large \bf   Geometry and observables in (2+1)-gravity}

\baselineskip 16pt

\vspace{.7cm}
{\large C.~Meusburger\footnote{\tt  catherine.meusburger@uni-hamburg.de}

Department Mathematik\\
Universit\"at Hamburg\\
Bundesstra\ss e 55, D-20146 Hamburg, Germany} \\

\vspace{0.5cm}

{28 January 2010}

\end{center}

\begin{abstract}

\noindent We review  the geometrical properties of vacuum spacetimes in (2+1)-gravity with vanishing cosmological constant. We explain how these spacetimes are characterised as quotients of their universal cover by holonomies.
 We explain how this description can be used to clarify the geometrical interpretation of the fundamental physical variables of the theory, holonomies and Wilson loops.  In particular,  we discuss the role of  Wilson loop observables as the generators of the two fundamental  transformations that change the geometry of (2+1)-spacetimes, grafting and earthquake. We explain how these variables can be determined from realistic measurements by an observer in the spacetime.
\end{abstract}

\section{Introduction}
\label{intro}

Gravity in (2+1) dimensions serves as a toy model that allows one to investigate conceptual questions of (quantum) gravity in a mathematically tractable theory. As many questions related to time, light and causality cannot be addressed in a Euclidean framework, 
(2+1)-spacetimes of Lorentzian signature are particularly relevant in this context. 
These spacetimes have a rich geometry, for an overview see \cite{Carlipbook,bb}, and realistic physical features such as initial (big bang) singularities and  cosmological time functions.

In order to obtain  interesting physics from these models, one needs to relate
 the variables which are used in the parametrisation of the phase space and in the quantisation of the theory  to the geometry  of the associated  spacetime. This has proven challenging even in the classical theory. It is not obvious what is the geometrical interpretation  of the fundamental 
  diffeomorphism invariant observables  (Wilson loops) and the associated phase space transformations they generate via the Poisson bracket. 
It  is  also   not apparent  how  these variables could be determined through  measurements  performed by an observer in the spacetime.

In this paper, we summarise the results of \cite{ich,ich3,icht} which provide an answer to these questions. In Section \ref{backgr} we review Lorentzian (2+1)-spacetimes without matter and cosmological constant. We discuss their geometrical properties and their description as quotients of their universal cover. Section \ref{earthgr}  introduces  the two fundamental  transformations that change the geometry of these spacetimes, grafting and earthquake along closed spacelike geodesics \cite{mess}.
 In Section \ref{geomobs} we explain how these two geometry changing transformations are generated via the Poisson bracket by the two fundamental  Wilson loop observables associated to the geodesic. This provides a physical interpretation of Wilson loop observables. 
 In Section \ref{lightmeas} we   discuss how the fundamental variables of the theory (holonomies and Wilson loops) 
  can  be determined from measurements by observers in the spacetime.

\section{Vacuum spacetimes in (2+1)-gravity}
\label{backgr}

\subsection{Gravity in (2+1)-dimensions}

As the Ricci curvature on a three-dimensional manifold determines its sectional curvature, vacuum solutions of the three-dimensional 
Einstein equations without cosmological constant are flat. The theory has no {\em local} gravitational degrees of freedom, and  its solutions are locally isometric to three-dimensional Minkowski space $\mathbb M^3$.  The solutions considered in this paper  are obtained as quotients 
 of  regions in  Minkowski space $\mathbb  M^3$ by discrete subgroups of its isometry group.  This subgroup encodes the {\em global} physical degrees of freedom  of the theory which arise from the non-trivial topology of the spacetimes.
 
 The  group of orientation and time orientation preserving isometries of  Minkowski space is the three-dimensional Poincar\'e group  $P_3=SO^+(2,1)\ltimes \mathbb R^3$, which is  the semidirect product of the proper orthochronous  Lorentz group $SO^+(2,1)$ in three dimensions with the translation group $\mathbb R^3$. In the following, we parametrise elements of $P_3$ as $(u,\ba)$ with $u\in SO^+(2,1)$ and $\ba\in\RR^3$. The group multiplication law of $P_3$ then takes the form
\begin{align}
\label{p3mult}
(u_1,\ba_1)\cdot (u_2,\ba_2)=(u_1u_2,\ba_1+u_1\ba_2).
\end{align}
We denote by $\exp: \mathfrak{so}(2,1)\rightarrow SO^+(2,1)$ the exponential map and introduce a basis $\{J_a\}_{a=0,1,2}$
of $\mathfrak {so}(2,1)$  in which the Lie bracket is given by
$[J_a,J_b]=\epsilon_{abc} J^c$. Here and in the following
 indices are raised and lowered with the Minkowski metric $\eta=\text{diag}(-1,1,1)$ and $\epsilon$ is the totally antisymmetric tensor  in three indices with $\epsilon_{012}=1$.

\subsection{Classification and geometry of  (2+1)-spacetimes without matter}

Under certain additional assumptions, the absence of local gravitational degrees of freedom allows one to  classify  the  solutions of the (2+1)-dimensional vacuum Einstein equations.
Such a classification has been achieved for {\em maximally globally hyperbolic} (MGH) flat (2+1)-spacetimes $M$ with geodesically {\em  complete} Cauchy surfaces \cite{mess,npm}.
The condition of global hyperbolicity  is imposed to exclude spacetimes with bad causality properties.  It ensures the absence of closed, timelike curves and  the existence of a  Cauchy surface $S$, a spacelike surface that every inextensible causal curve intersects exactly once. It also implies that $M$ is of topology $M\approx\RR\times S$ and $\pi_1(M)\cong \pi_1(S)$. The maximality is a technical condition that avoids over-counting spacetimes, and the completeness condition  excludes Cauchy surfaces with singularities.

\begin{figure}[h]
\centering
a)\includegraphics[scale=0.2]{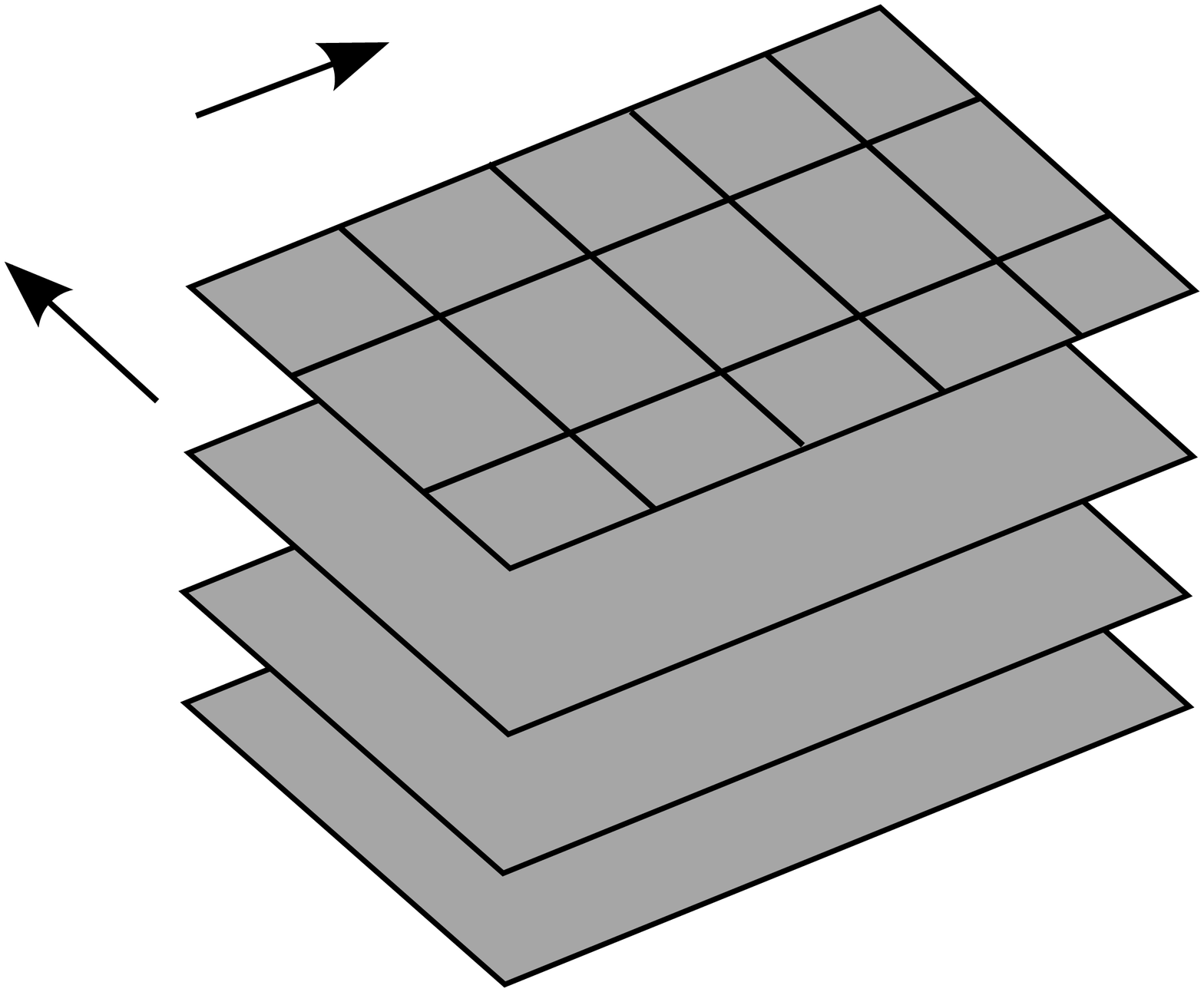}
b)\includegraphics[scale=0.2]{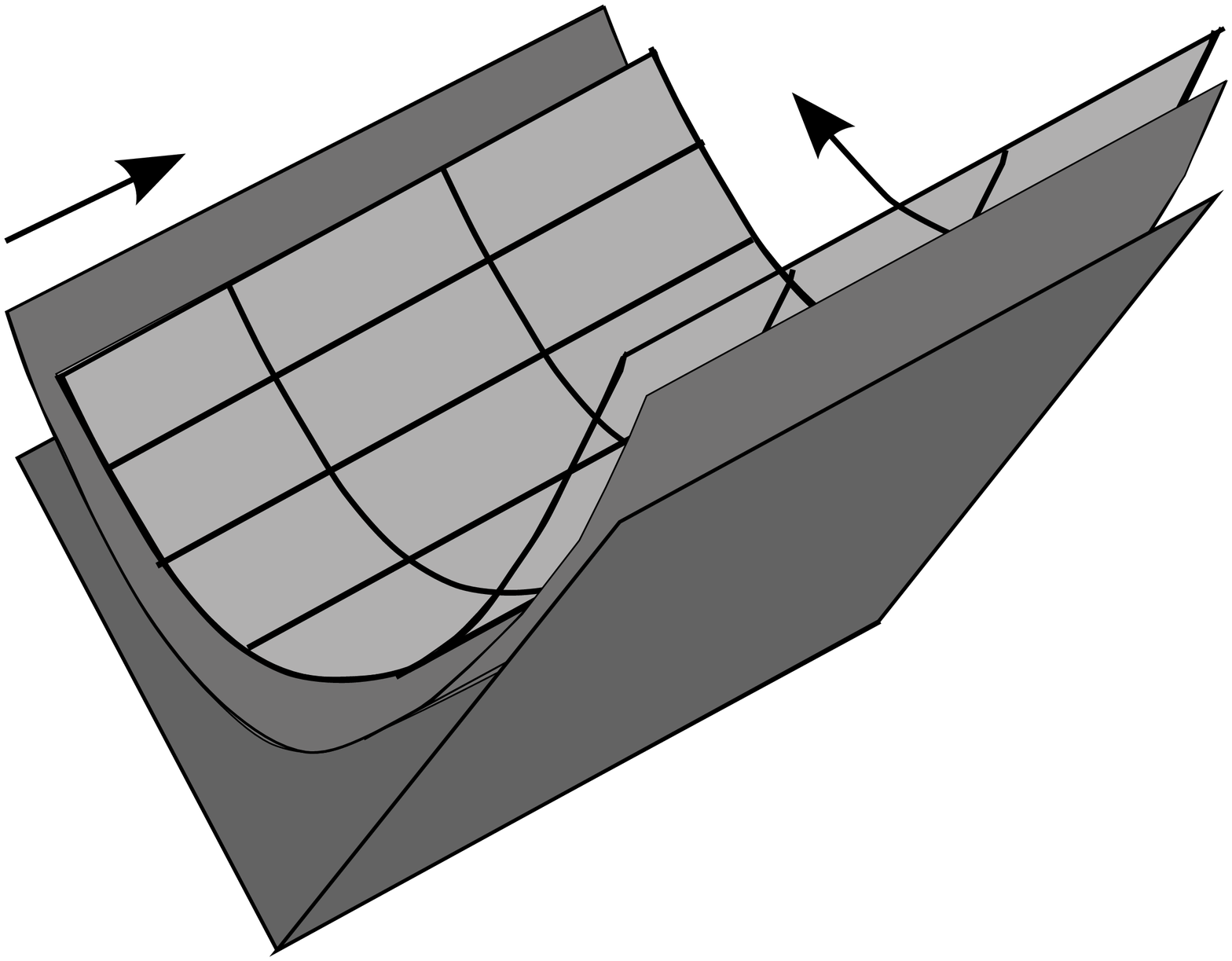}
\caption{\small{Universal cover  for the torus universe. Case a): Static torus universe; $\tilde M=\mathbb M^3$. Case b:) Evolving torus universe; the universal cover $\tilde M$ is the future of a spacelike geodesic in $\mathbb M^3$. The grids and arrows indicate the action of $\pi_1(M)$ on $\tilde M$.}}
\label{torusdomain}
\end{figure}

 It is shown in \cite{barbot} that the universal cover  $\tilde M$ of a MGH flat (2+1)-spacetime $M$ with a complete Cauchy surface  can be identified with a domain of dependence in Minkowski space, i.~e.~the future of a set of points in $\mathbb M^3$.
One distinguishes the following four cases:
\begin{description}
\item [Case 1: $\pi_1(M)=\{1\}$.]  In this case $\tilde M=M$ is either the entire Minkowski space $\mathbb M^3$ or the future  of a lightlike plane $P$  or the intersection  of the future  of a lightlike plane $P$ with the past  of a lightlike plane $Q$ parallel to $P$.  
\item [Case 2: $\pi_1(M)=\mathbb Z$ (Cylinder universe).] In this case,  the Cauchy surface has the topology of a cylinder, and the domains are the same as in case 1. The elements of $\pi_1(M)$ act by spacelike translations.
\item [Case 3: $\pi_1(M)=\mathbb Z\oplus \mathbb Z$ (Torus universe).] In this case, the domain is either Minkowski space $\tilde M=\mathbb M^3$, and both generators of $\pi_1(M)$ act by spacelike translations as shown in Figure \ref{torusdomain} a) or $\tilde M$ is the future of a spacelike line  as shown in Figure \ref{torusdomain} b). In that case, one of the generators acts by spatial translations in the direction of the geodesic's tangent vector, while the other is a boost  with this tangent vector as its axis. 
\item [Case 4: $\pi_1(M)\neq \{1\},\mathbb Z,\mathbb Z\oplus\mathbb Z$ (Higher genus case).] In this case, $\tilde M$ is the future of set of points  different from a geodesic line. The Lorentzian part of the holonomy $h: \pi_1(M)\rightarrow SO^+(2,1)$ defines a faithful and discrete representation of $\pi_1(M)$.
\end{description}

As case 1 and 2 have too few degrees of freedom and case 3  has been investigated extensively  (for an overview see \cite{Carlipbook}, for a discussion of  observers and geometry  \cite{gua}), we focus on  case 4 in the following. Moreover, we restrict attention to the simplest case, namely to spacetimes which have a {\em compact}, orientable Cauchy surface $S$ of genus $g\geq 2$. 

In this case, the universal cover $\tilde M$ is either the future of a point $\bp\in\mathbb M^3$ or of a  graph $G\subset \mathbb M^3$ \cite{bb,mess, barbot}. Both, $M$ and  $\tilde M$, are future complete and  have an initial singularity. Moreover, it is shown in \cite{bb,mess,bg} that $M$ and $\tilde M$ are equipped with a cosmological time function, which gives the geodesic distance of points in $M,\tilde M$ from the initial singularity, and that they are foliated by surfaces $M_T, \tilde M_T$ of constant cosmological time.

\begin{figure}[h]
\centering
a)\!\!\!\! \!\!\!\! \!\!\!\! \includegraphics[scale=0.15]{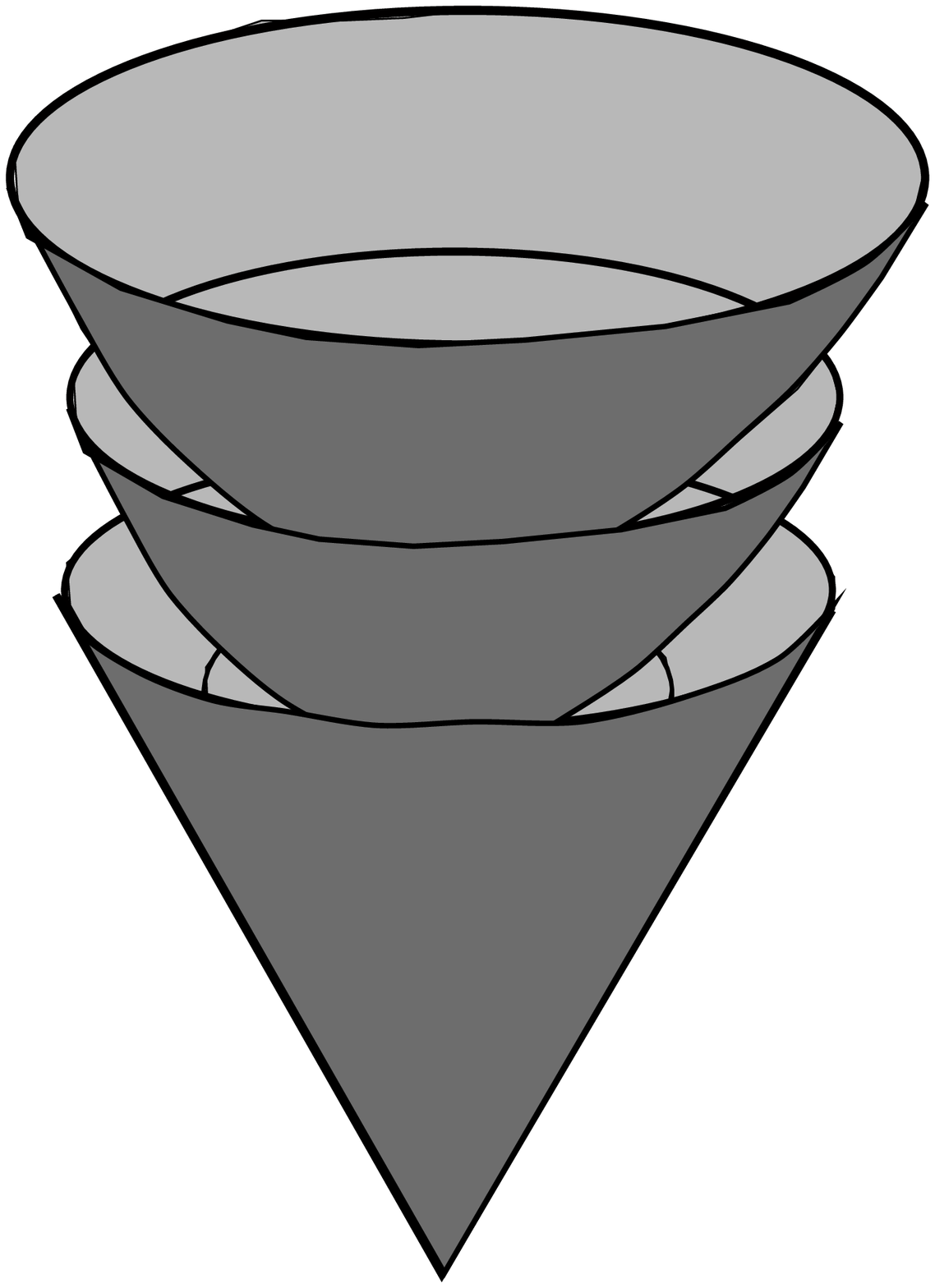}
b) \includegraphics[scale=0.15]{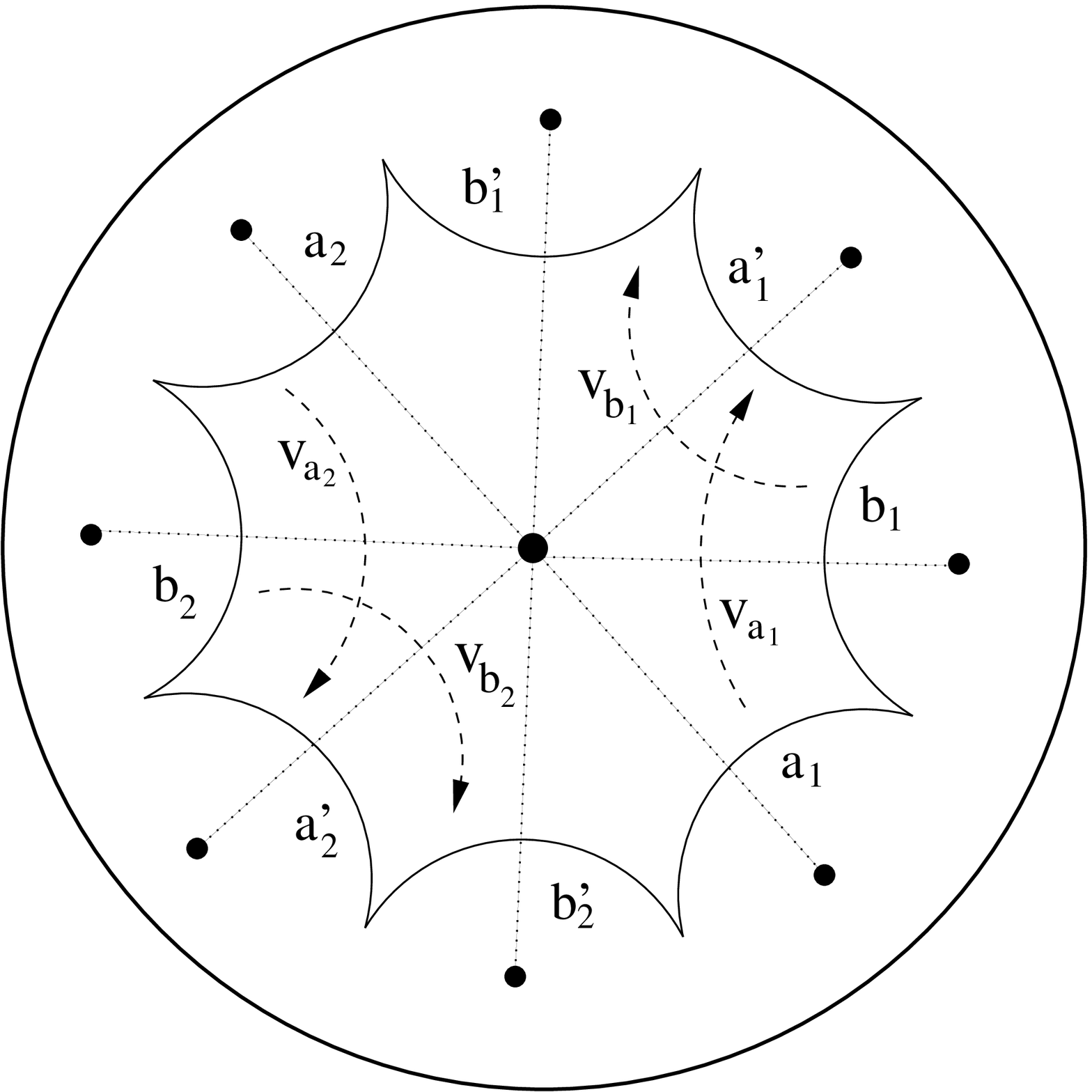}
c)\includegraphics[scale=0.15]{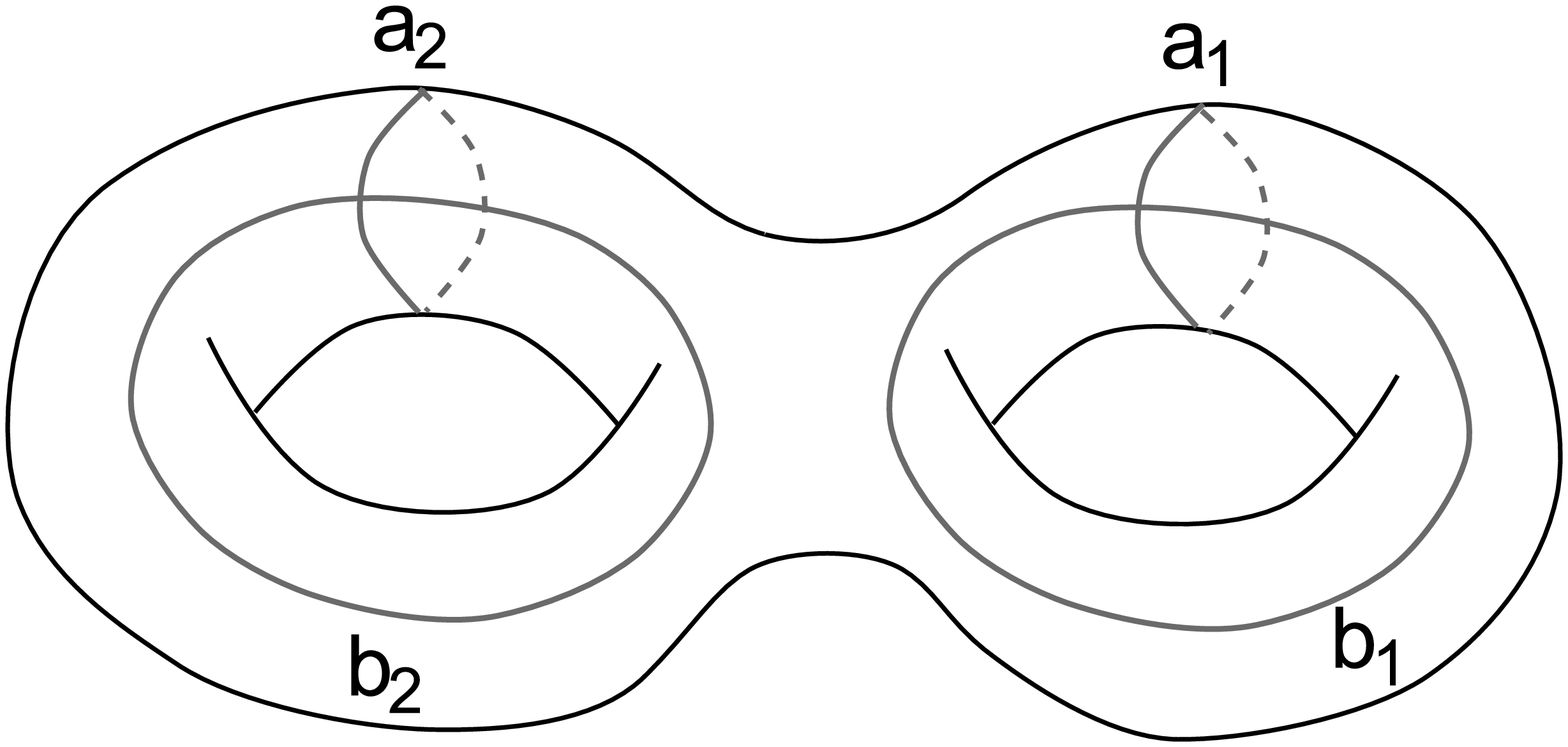}
\caption{\small{a) Foliation of the lightcone by hyperboloids. b) Identification of points in $\tilde M_T\cong \hyp$ for a genus $g=2$-surface. c) Resulting genus $g=2$ Riemann surface.}}
\label{lightcone}
\end{figure}

The fundamental group $\pi_1(M)$ acts on $\tilde M$ freely and properly discontinuously in such a way that each constant cosmological time surface $\tilde M_T$ is preserved. This group action is given by the {\em holonomies}, a group homomorphism 
$h:\pi_1(M)\rightarrow P_3$ into the isometry group of Minkowski space. 
The Lorentzian components $v_\lambda$ of  the holonomies $h(\lambda)=(v_\lambda,\ba_\lambda)$, $\lambda\in\pi_1(M)$, define a cocompact Fuchsian group  of genus $g$. This is a discrete subgroup $\Gamma\subset SO^+(2,1)$ of the three-dimensional Lorentz group with $2g$ generators and a single defining relation, which is isomorphic to $\pi_1(M)$
\begin{align}
\Gamma=\langle v_{a_1}, v_{b_1},..., v_{a_1}, v_{b_1}\;|\; [v_{B_g}, v_{A_g}]\cdots[v_{B_1}, v_{A_1}]=1\rangle\qquad[v_{b_i}, v_{a_i}]=v_{b_i} v_{a_i}v_{b_i}^\inv v_{a_i}^\inv.
\end{align}
The spacetime $M$ is obtained as the quotient of $\tilde M$ by this group action, i.~e.~by identifying on each surface $\tilde M_T$ the points that are mapped into each other by the holonomies
\begin{align}
M=\bigcup_{T\in\RR^+} M_T=\bigcup_{T\in\RR^+} \tilde M_T/\pi_1(M)_h.
\end{align}
The simplest example are the conformally static spacetimes for which the universal cover is a lightcone, i.e. the future of a point $\bp\in\mathbb M^3$. In this case, the constant cosmological time surfaces  are hyperboloids $\tilde M_T=T\cdot \hyp$ as depicted in Figure \ref{lightcone} a).
As the action of  $\pi_1(M)$ on $\tilde M$  preserves these hyperboloids,  its translational component is trivial, i.~e.~given by conjugation
\begin{align}
h(\lambda)=(1,-\bp)\cdot(v_\lambda,0)\cdot (1,\bp)\qquad\forall \lambda\in\pi_1(M).
\end{align}
The action of $\pi_1(M)$ on the surfaces $\tilde M_T$ coincides with the canonical action of the associated Fuchsian group $\Gamma$ on two-dimensional hyperbolic space $\hyp$. 
This group action induces a tessellation of each constant cosmological time surface by geodesic arc $4g$-gons which are mapped into each other by the elements of $\Gamma$. The elements of $\Gamma$ identify the sides of these geodesic arc $4g$-gons pairwise as indicated in Figure \ref{lightcone} b). 

The constant cosmological time surfaces $M_T$ are obtained by identifying the points on the hyperboloids $\tilde M_T$ which are related by this action of $\Gamma$. This amounts to gluing the sides of each polygon pairwise as shown in Figure \ref{lightcone} b).  
Each constant cosmological time surface $M_T$  is thus a copy of the same Riemann surface  as in  Figure \ref{lightcone} c), rescaled by the cosmological time $M_T=T\cdot \Sigma_\Gamma=T\cdot \hyp/\Gamma$ and thus equipped with a  metric of constant curvature $-1/T^2$. As the geometry of the constant cosmological time surfaces $M_T$ evolves with $T$ only by a rescaling, these spacetimes are called conformally static.

\subsection{Phase space}

It is shown in \cite{mess}, see also \cite{npm}, that flat maximally globally hyperbolic (2+1)-spacetimes of genus $g\geq 2$ and their universal covers are characterised uniquely by the holonomies $h:\pi_1(M)\rightarrow P_3$.  Two spacetimes are isometric if and only if the associated holonomies are related by global conjugation with a constant element of $P_3$. Moreover, each group homomorphism $h: \pi_1(M)\rightarrow P_3$ whose Lorentzian component defines a cocompact Fuchsian group of genus $g$ gives rise to such a spacetime. This implies that the physical phase space of the theory, the space of solutions of the (2+1)-Einstein equations modulo diffeomorphisms,  coincides with  the cotangent bundle $T^*\tau_g$  of Teichm\"uller space $\tau_g$ and is given by
\begin{align}
\label{phsp}
\mathcal P=\text{Hom}_0(\pi_1(S), P_3)/P_3=T^*\tau_g,
\end{align}
where the index $0$ indicates the restriction on the Lorentzian component of the holonomy. 
The phase space is equipped with a canonical symplectic structure, the Goldman bracket \cite{gol}.

The physical (diffeomorphism invariant) observables of the theory are  functions on $\mathcal P$ or, equivalently, functions of the holonomies $h(\lambda)$, $\lambda\in\pi_1(M)$, which are invariant under simultaneous  conjugation with $P_3$. A complete set of such observables is given by the Wilson loops.
In the case at hand,  there are two canonical Wilson loop observables associated to each element $\lambda\in\pi_1(M)$. 
The first, in the following referred to as {\em mass} and denoted $m_\lambda$, depends only on the Lorentzian component of the associated holonomy, while the second, the {\em spin} $s_\lambda$, involves both the Lorentzian and the translational component
\begin{align}\label{ms}
m_\lambda=\sqrt{\bp_\lambda^2}\qquad s_\lambda=\hat \bp_\lambda\cdot \ba_\lambda\qquad\text{where}\qquad h(\lambda)=(\exp(p^a_\lambda J_a), \ba_\lambda), \;\hat\bp_\lambda=\bp_\lambda/m_\lambda
\end{align}
These Wilson loop observables are closely related to the two Casimir operators of the three-dimensional Poincar\'e algebra and play a fundamental role in the quantisation of the theory.

\section{Construction of spacetimes via earthquake and grafting}
\label{earthgr}

Given a conformally static spacetime $M$ and a closed, simple geodesic $\lambda$ on a constant cosmological time surface $M_T$, there are two canonical ways of changing the geometry of $M$ 
 \cite{th2,mcmull}: {\em grafting} and {\em earthquake}.  
The ingredients in the construction are a cocompact Fuchsian group $\Gamma$, given by the Lorentzian component of the  holonomies, and a closed, simple geodesic\footnote{Note that grafting and earthquake are defined in a more general context, namely for measured geodesic laminations on a Riemann surface. For our purposes it is sufficient to consider the simplest case, in which the measured geodesic lamination is a geodesic with weight $w\in \RR^+$.} on the associated Riemann surface $\hyp/\Gamma$ with a weight $w\in\mathbb R^+$.

\begin{figure}[h]
\centering
a) \includegraphics[scale=0.25]{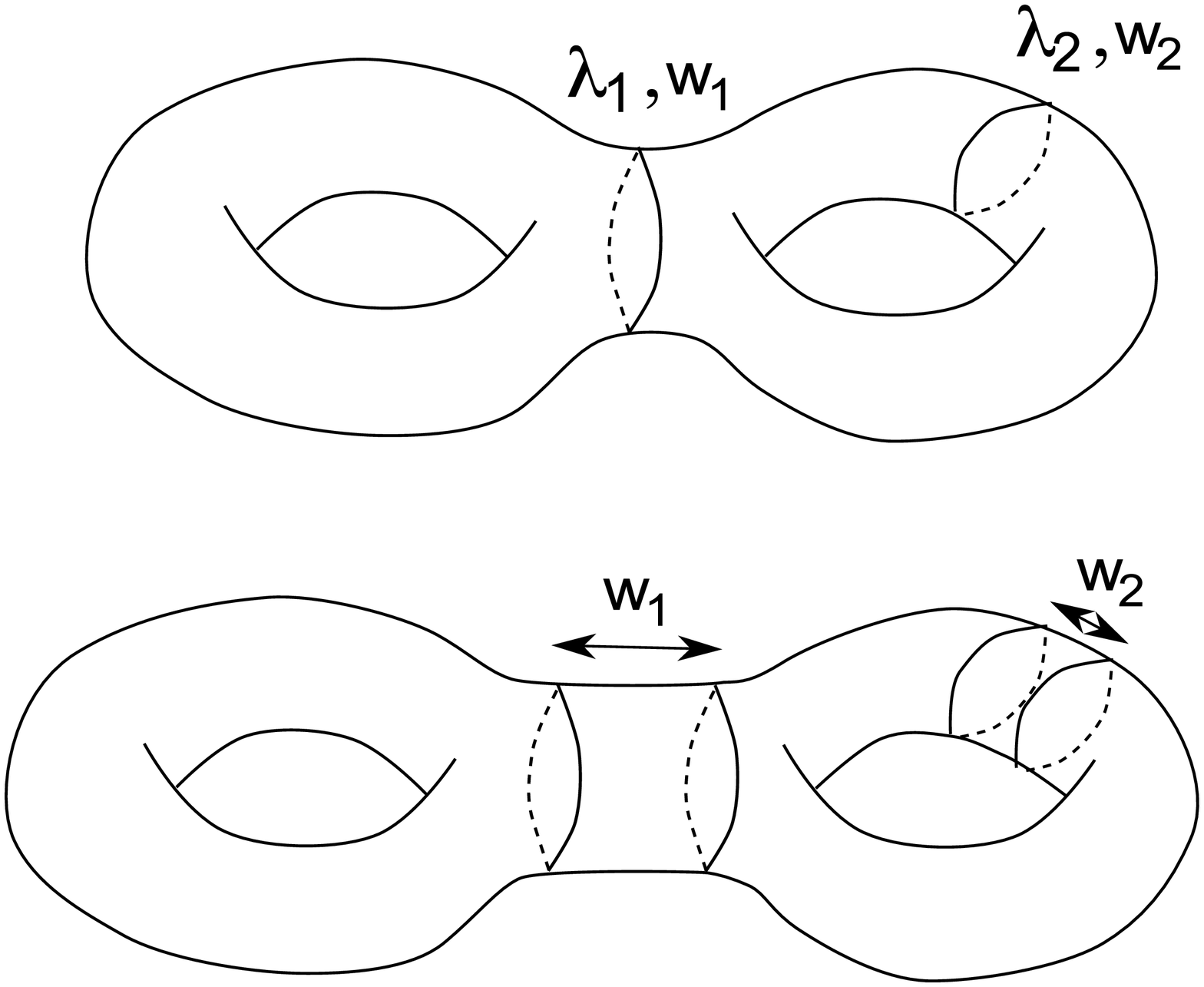}
b)\includegraphics[scale=0.25]{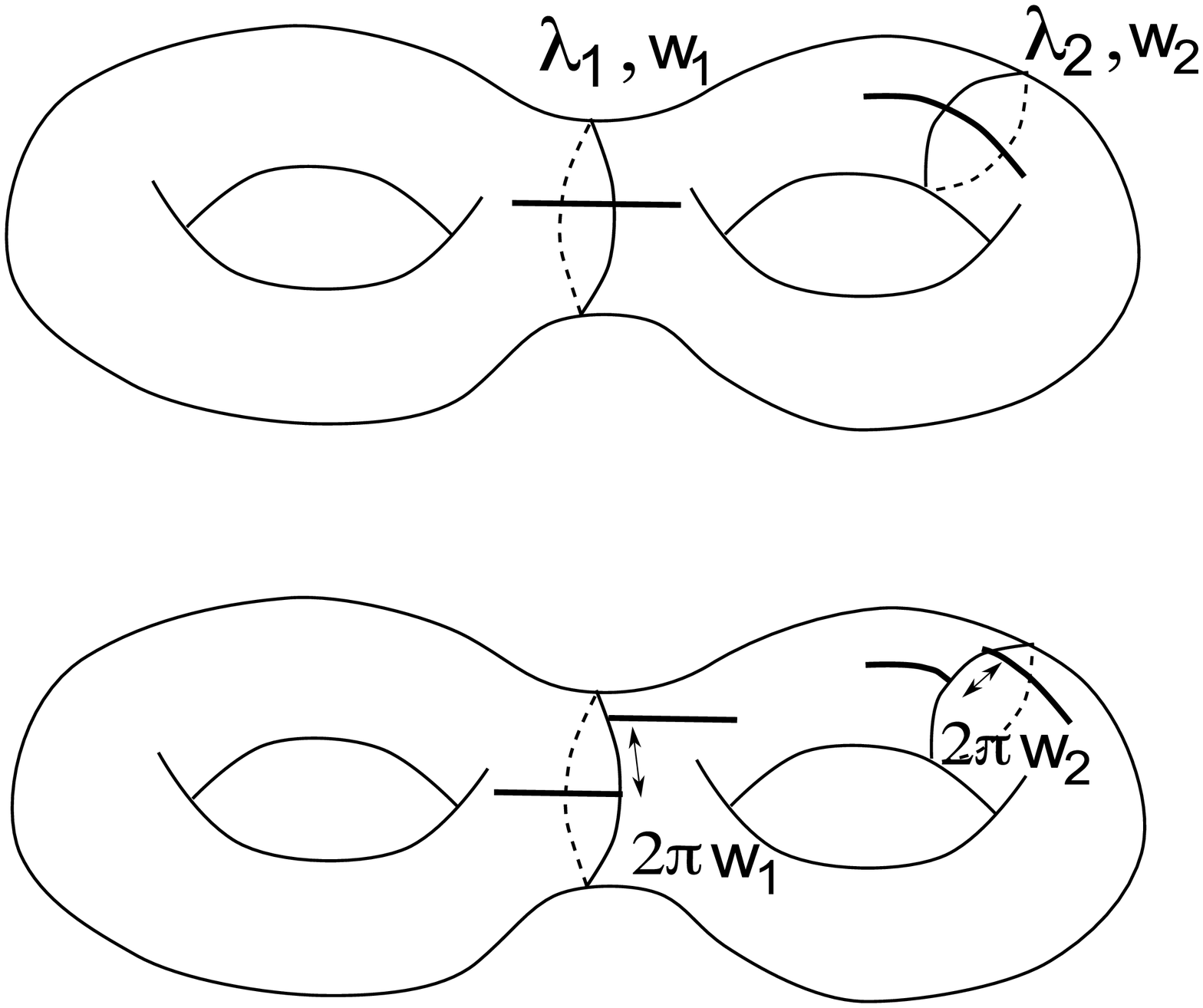}
\caption{\small{Grafting (a) and earthquake (b) along  weighted geodesics on a genus two surface.}}
\label{graftsurf}
\end{figure}

Schematically, grafting amounts to cutting a Riemann surface along the geodesic $\lambda$ and inserting a cylinder of width $w$, as shown in Figure \ref{graftsurf} a). The earthquake corresponds to cutting the Riemann surface along the geodesic  and rotating the edges of the cut against each other by an angle $2\pi w$ as shown in Figure \ref{graftsurf} b).

\subsection{Deformation of the universal cover}
\begin{figure}[h]
\centering
a)\includegraphics[scale=0.3]{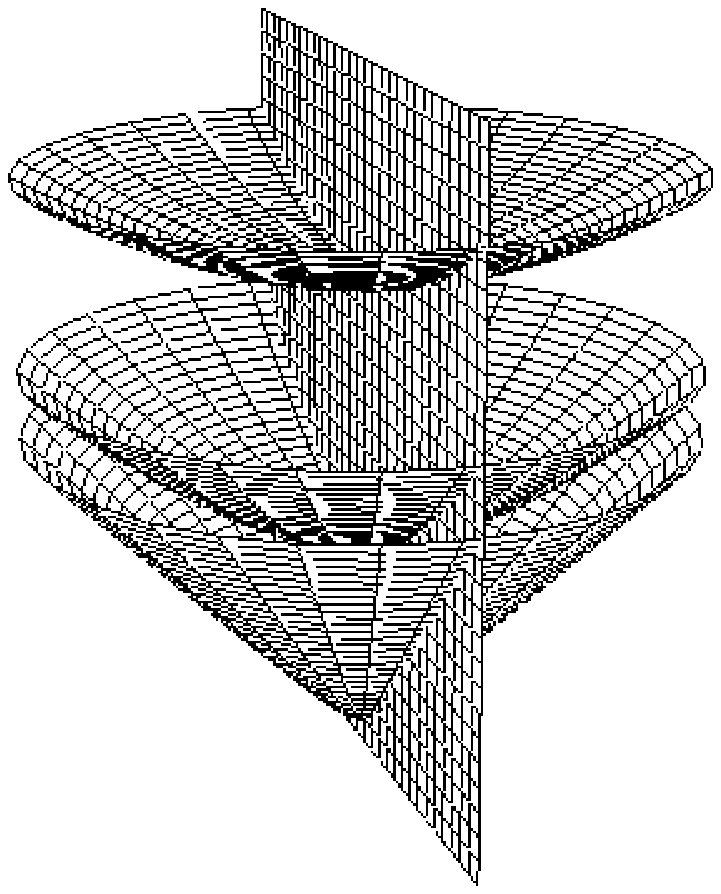}
b)\includegraphics[scale=0.3]{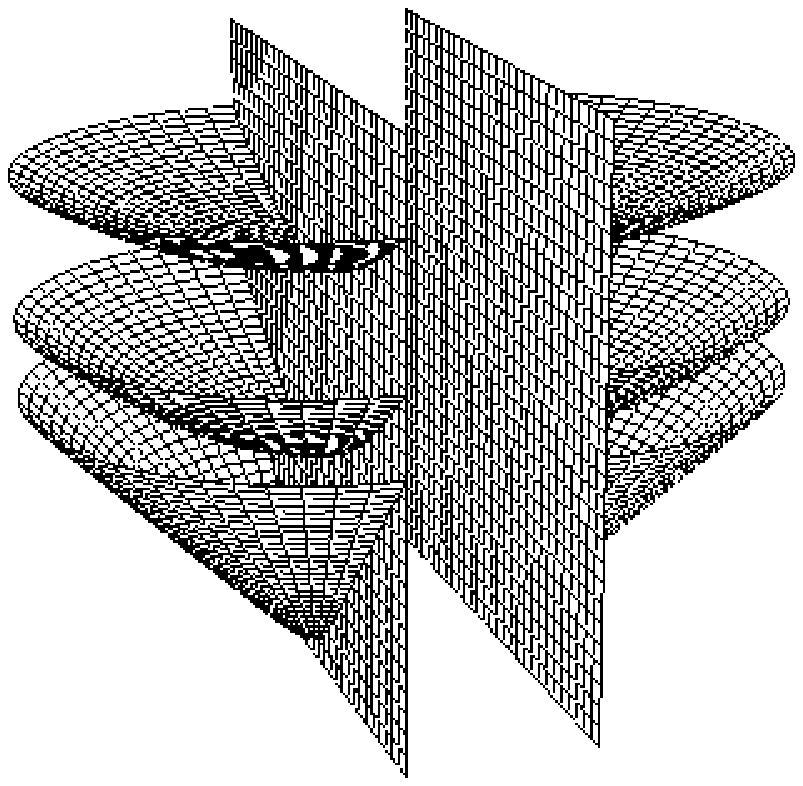}
c)\includegraphics[scale=0.3]{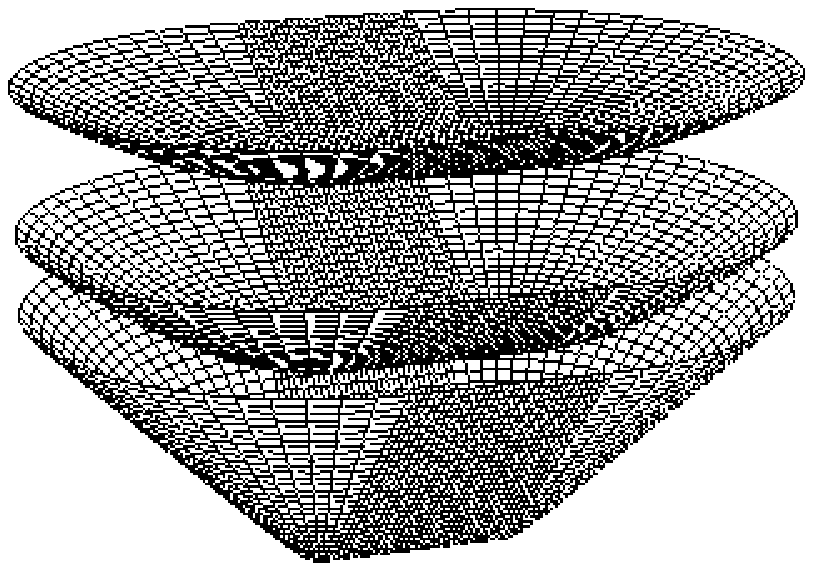}
d)\includegraphics[scale=.3]{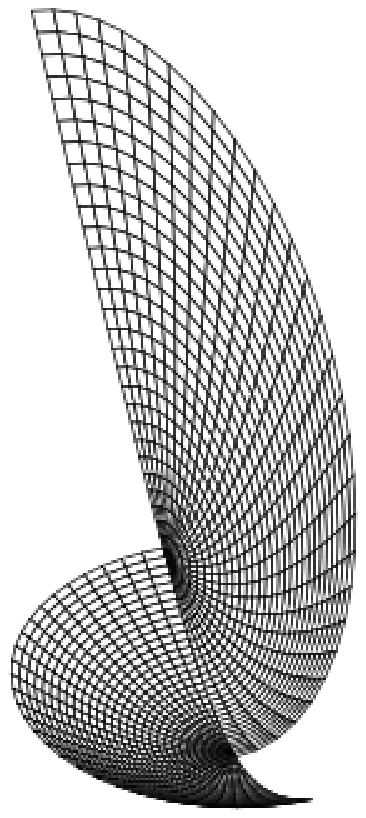}
\caption{\small{Grafting and earthquake construction  in the lightcone for a single geodesic:\newline
a) Intersection of lightcone with the plane defined by the geodesic\newline
b) Cutting the lightcone and shifting the pieces in the grafting construction\newline
c) Deformed domain for the grafting construction.\newline
d) Earthquake on a constant cosmological time surface
}}
\label{grafting}
\end{figure}

In the application to conformally static (2+1)-spacetimes, grafting and earthquake are performed simultaneously on each constant cosmological time surface $M_T=T\cdot \hyp/\Gamma$ in such a way that the weight is the same for all surfaces.  The construction can be implemented in the universal cover $\tilde M$, i.~e.~a lightcone in Minkowski space. For this, one lifts the geodesic on $\hyp/\Gamma$  to a multicurve, an infinite set of non-intersecting, weighted geodesics on each hyperboloid $\tilde M_T=T\cdot \hyp$. These geodesics are obtained  by taking one lift of the geodesic to $\tilde M_T$ and acting on it with the Fuchsian group $\Gamma$. They are given as the intersection of the hyperboloids with planes through the tip of the lightcone as shown in Figure \ref{grafting} a).

In the grafting construction, one selects a basepoint  $\bq\in \tilde M$ outside  these planes and cuts the lightcone along all planes defined by geodesics in the multicurve.  
One then shifts the pieces that do not contain the basepoint $\bq$ away from $\bq$, in the direction of the plane's normal vector by a distance given by the weight as shown in Figure \ref{grafting} b).  The translated pieces of the lightcone are then joint by
 straight lines as shown in Figure \ref{grafting} c). This yields a deformed domain $\tilde M$ which is no longer the future of a point but the future of a graph. The surfaces $\tilde M_T$ of constant cosmological time are surfaces of constant geodesic distance from this graph. They are deformed hyperboloids  with strips glued in along each geodesic in the multicurve as shown in Figure \ref{grafting} c). 

In the earthquake construction, one proceeds analogously. As in the case of grafting, one selects a basepoint and cuts the lightcone along all planes defined by geodesics in the multicurve. One then applies  Lorentz boosts to the pieces that do not contain the basepoint. These boosts are chosen such that they
 preserve the associated planes and have the weight $w\in\RR^+$ as their rapidity.
 Their action on the hyperboloids  is depicted in Figure \ref{grafting} d). 
 They preserve the domain $\tilde M$ and its foliation by constant cosmological time surfaces $\tilde M_T$. The resulting spacetime remains conformally static.

\subsection{Action of grafting and earthquake on the holonomies}

The action of the fundamental group $\pi_1(M)$ on the deformed domains via the holonomies $h: \pi_1(M)\rightarrow P_3$ is defined in such a way that it identifies two points in the deformed domain if and only if the corresponding points in the original lightcone were identified.
In the case of a grafted spacetime, this implies that the holonomies acquire a non-trivial translational component that takes into account the translations  and grafted strips.  Grafting therefore does not affect the Lorentzian components of the holonomies, which determine the Fuchsian group $\Gamma$, but changes their translational components.
 For the spacetimes deformed by  earthquakes, the translational components of the holonomies remain trivial, but the Lorentzian components  change to take into account the Lorentz boosts applied to different pieces of the lightcone.

The transformation of the holonomies associated with grafting and earthquakes can be determined explicitly. As the holonomies characterise the spacetime uniquely, it follows that grafting and earthquakes induce transformations on the  phase space $\mathcal P$ \eqref{phsp}. These phase space transformations 
 are given explicitly in \cite{bb, ich,ich3}.

\subsection{Spacetimes}

\begin{figure}[h]
\centering
 \includegraphics[scale=0.3]{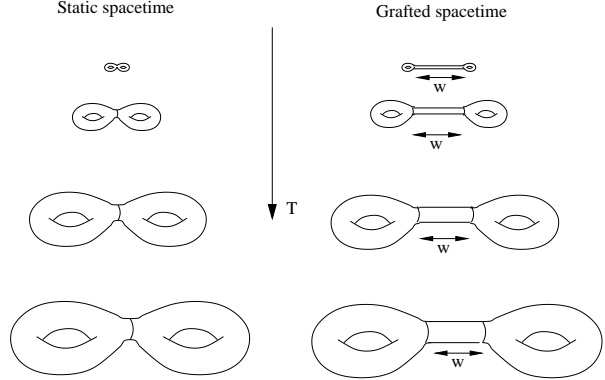}
\caption{\small{Evolution of the surfaces  $M_T$ in a conformally static and grafted spacetime.}}
\label{grtimevol}
\end{figure}

The deformed spacetimes obtained via the grafting and earthquake construction are given as the quotients of the deformed domains by the action of the fundamental group  via the  holonomies $h:\pi_1(M)\rightarrow P_3$. 
In the case of a grafted spacetime, one finds that the resulting spacetimes are no longer conformally static, but show a non-trivial evolution with the cosmological time as indicated in Figure \ref{grtimevol}. While the constant cosmological time surfaces of a conformally static spacetime are all copies of the same Riemann surface $\Sigma=\hyp/\Gamma$, rescaled by the cosmological time, this is no longer true for the grafted spacetimes. Although the part outside the grafted strips is rescaled with the cosmological time, 
 the width of the grafted strips remains constant.   This implies that the impact of grafting is maximal near the initial singularity and vanishes in the limit $T\rightarrow \infty$ as shown in Figure \ref{grtimevol}. 
 
 The grafting construction thus maps the conformally static spacetime associated with a cocompact Fuchsian group $\Gamma$ to an evolving spacetime associated with $\Gamma$, which approaches the conformally static spacetime in the limit $T\rightarrow\infty$.
In contrast, the earthquake preserves the constant cosmological time surfaces and affects only the Lorentzian component of the holonomies. It maps a conformally static spacetime associated with a Fuchsian group $\Gamma$  to a conformally static
spacetime associated with a non-equivalent Fuchsian group $\Gamma'$.

\section{Geometry transformations and observables }
\label{geomobs}

As the  fundamental diffeomorphism invariant observables of the theory which  characterise the geometry of the spacetimes  and parametrise the  phase space $\mathcal P$ \eqref{phsp}, the Wilson loop observables  play an important role in quantisation. 
This makes it natural to investigate the geometrical interpretation of the   mass and spin observables \eqref{ms}. It also  raises the question how the phase space transformations that are generated by these observables  via the Poisson bracket affect the geometry of the spacetime.  

\begin{theorem} \cite{ich,ich3}   Let $M\approx\mathbb R\times S$ be a a conformally static vacuum spacetime of genus $g\geq 2$ and $\lambda$ a closed, simple geodesic on a constant cosmological time surface $M_T$  with associated weight $w$. 
Then the two fundamental Wilson loop observables, the mass $m_\lambda$ and spin $s_\lambda$,   associated to  $\lambda$ generate via the Poisson bracket  the phase space transformations $\text{Gr}_\lambda^w, E_\lambda^w: \mathcal P\rightarrow \mathcal P$ associated with geometry change through grafting and earthquake along $\lambda$
\begin{align}
\{m_\lambda, f\}=\frac d {dw}\bigg|_{w=0}f\circ \text{Gr}_\lambda^w\qquad\qquad \{s_\lambda, f\}=\frac d {dw}\bigg|_{w=0}f\circ \text{E}_\lambda^w\qquad \forall f\in\cif(\mathcal P).
\end{align}
\end{theorem}
This theorem confirms that the Wilson loop observables  which are fundamental in the quantisation of the theory also play a distinguished role with respect to spacetime   geometry:  they generate via the Poisson bracket  the two fundamental  transformations which change the geometry of the spacetime.  In particular, this  implies  that the grafting and earthquake transformations are Poisson isomorphisms (canonical transformations). 

It also provides one with a  clear physical  interpretation of the two fundamental Wilson loop observables which justifies the names ``mass" and ``spin". The mass observable generates grafting, the spin observable earthquakes, which can be viewed, respectively, as a translation and a rotation associated with a geodesic. 
In analogy with classical mechanics, where momenta generate translations and angular momenta rotations, the mass and spin observable can thus be interpreted as a momentum  and an angular momentum  of  a geodesic.

\section{Measurements associated with returning lightrays}
\label{lightmeas}

Although the geometry of Lorentzian vacuum (2+1)-spacetimes is well-understood, it has been difficult  to obtain interesting physics from the theory. This is due to the fact that
it is not readily apparent how the variables which parametrise the phase space, holonomies and Wilson loops,  could be measured by an observer in the spacetime. 
Moreover, it is not obvious  what physically meaningful measurement an observer could make in an empty spacetime and how such measurements would allow him to distinguish the different spacetimes that are all locally isometric to Minkowski space.

This problem is addressed in \cite{icht}, where it is shown that an observer can determine the geometry of such spacetimes by sending out lightrays which return to him at a later time.  
The observer can then measure the return time, the eigentime elapsed between the emission of the lightray and its return. He can determine into which directions the light needs to be sent in order to return, and he can compare the frequencies of the emitted and returning lightray.

\begin{figure}[h]
\centering
 \includegraphics[scale=0.25]{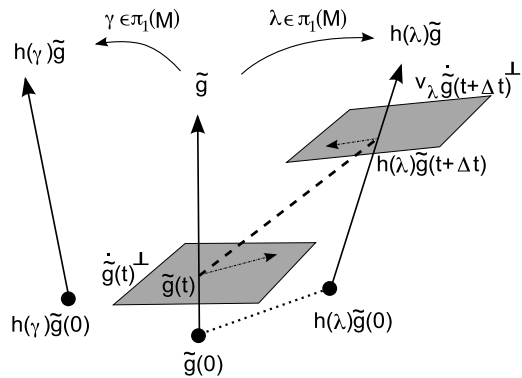}
\caption{\small{Returning lightray in the universal cover $\tilde M\subset\mathbb M^3$}}
\label{cover}
\end{figure}

To obtain explicit expressions for these measurements in terms of the holonomy variables, it is advantageous to work in the universal cover $\tilde M\subset\mathbb M^3$. 
We start by explaining how the relevant concepts (observers, lightrays, returning lightrays) are implemented in this description. 
By definition, an observer in free fall in $M$ is characterised by a timelike, future oriented geodesic  $g:[0,\infty)\rightarrow M$. This corresponds to a $\pi_1(M)$-equivalence class of timelike, future directed geodesics in the universal cover $\tilde M$. This equivalence class of geodesics is obtained   by taking one lift $\tilde g:[0,\infty)\rightarrow \tilde M$ of $g$ to $\tilde M$ and constructing its images $h(\lambda)\tilde g$, $\lambda\in\pi_1(M)$  under the holonomies as shown in Figure \ref{cover}. 

A lightray is characterised by  lightlike geodesic in $M$ or, equivalently, the $\pi_1(M)$-equivalence class of lightlike, future oriented geodesics in $\tilde M$ obtained by taking one lift and acting on it with the holonomies. A returning lightray with respect to an observer with wordline $g:[0,\infty)\rightarrow M$ is a lightray in $M$ that intersects the worldline of the observer twice. In the universal cover, such a returning lightray can be described as the $\pi_1(M)$-equivalence class of a  lightlike geodesic $\tilde M\subset\mathbb M^3$ that starts at one lift $\tilde g: [0,\infty)\rightarrow \tilde M$ of the observer's worldline and ends at one if its images $h(\lambda)\tilde g$, $\lambda\in\pi_1(M)$, as shown in Figure \ref{cover}.

As the universal cover  $\tilde M$ is a region in Minkowski space, the lifts of the observer's wordline and of lightrays are straight lines in Minkowski space with, respectively, timelike and lightlike tangent vectors.  The return time $\Delta t$  is then obtained by parametrising the observer's worldline $g:[0,\infty)\rightarrow M$ and its lifts $\tilde g:[0,\infty)\rightarrow \tilde M$ according to eigentime and  determining the unique positive solution of  the quadratic equation 
\begin{align}
\label{rett}\eta\big(\,h(\lambda)\tilde g(t+\Delta t)-\tilde g(t)\,,\,h(\lambda)\tilde g(t+\Delta t)-\tilde g(t)\,\big)=0.\end{align}
Returning lightrays are thus in one-to-one correspondence with elements of  $\pi_1(M)$. The directions into which the returning lightray associated with an element $\lambda\in\pi_1(M)$ is emitted and from which it returns are given by the projection of the lightlike vector $h(\lambda)\tilde g(t+\Delta t)-\tilde g(t)$ on the orthogonal complement of, respectively, the tangent vector  $\dot {\tilde g}(t)\in \RR^3$ and $v_\lambda \dot{\tilde g}(t+\Delta t)$, where $h(\lambda)=(v_\lambda,\ba_\lambda)$.
The relative frequencies of the emitted and returning lightray can be determined in analogy to the relativistic Doppler effect. One obtains
\begin{align}
\frac {f_r} {f_e}=\frac {\eta\big(\,h(\lambda)\tilde g(t+\Delta t)-\tilde g(t)\,,\, v_\lambda\dot {\tilde g}(t)\,\big)} {\eta\big(\,h(\lambda)\tilde g(t+\Delta t)-\tilde g(t)\,,\,\dot {\tilde g}(t)\,\big)}.
\end{align}
Using these definitions, one can derive explicit expressions for the return time, the directions and the relative frequency shift  \cite{icht}. 

\begin{theorem}\cite{icht}\label{retlr} Let $\tilde g: [0,\infty)\rightarrow \tilde M$ be the lift of the worldline of an observer in free fall in $M$, parametrised according to eigentime: 
\begin{align}\label{etime}
\tilde g(t)=t\bx+\bx_0\qquad\text{with}\qquad \bx\in\RR^3, \bx^2=-1, x^0>0, \bx_0\in \tilde M.
\end{align}
Consider the returning lightray associated with an element $\lambda\in\pi_1(M)$ that is emitted by the observer at eigentime $t$ and returns to him at $t+\Delta t$. Then the return time $\Delta t$
is given by
\begin{align}\label{retform}
\Delta t =(t+\sigma_\lambda)(\cosh \rho_\lambda -1)-\tau_\lambda+\sinh\rho_\lambda\sqrt{(t+\sigma_\lambda)^2+\nu_\lambda^2},
\end{align}
where $\rho_\lambda=-\bx\cdot (v_\lambda\bx)$ and
$h(\lambda)\bx_0-\bx_0=\sigma_\lambda(v_\lambda\bx-\bx) +\tau_\lambda v_\lambda\bx+\nu_\lambda\, \bx\wedge v_\lambda\bx$.
The direction into which the lightray is emitted is given by the spacelike unit vector \begin{align}\label{dirvec}
&\hat \bp^e_\lambda\!=\!\cos\phi_e \;\frac{v_\lambda\bx\!+\!(\bx\cdot v_\lambda\bx)\bx}{|v_\lambda\bx\!+\!(\bx\cdot v_\lambda\bx)\bx|}\!+\!\sin\phi_e\; \frac{\bx\wedge v_\lambda\bx}{|\bx\wedge v_\lambda\bx|}\;\;\text{with}\;\;
\tan \phi_e\!=\!\frac{\nu_\lambda}{\sinh\rho_\lambda\sqrt{(t\!+\!\sigma_\lambda)^2\!+\!\nu_\lambda^2}},
\end{align}
and the relative frequency $f_r/f_e$ of the lightray at its emission and return are given by
\begin{align}\label{frequshift}
f_r/f_e=\frac{\sqrt{(t+\sigma_\lambda)^2+\nu_\lambda^2}}{\cosh\rho_\lambda\sqrt{(t+\sigma_\lambda)^2+\nu_\lambda^2}+\sinh\rho_\lambda(t+\sigma_\lambda)}<1.
\end{align}
\end{theorem}
The physical interpretation of these measurements  is discussed in detail in \cite{icht}. It is shown there that for an observer  in a conformally static spacetime whose eigentime coincides with the cosmological time, the relative frequencies and the directions of emission do not vary with time and the return time is a linear function of the emission time. This is the case for a general observer in a general spacetime in the limit where the emission time $t$ and hence the cosmological time $T$  tend to infinity. In this limit the effect of grafting becomes negligible and the grafted spacetime approaches the associated conformally static spacetime.

The  measurements in Theorem \ref{retlr} thus allow an observer to draw conclusions about the geometry of the underlying spacetime. This raises the question if  an observer  can use them to  determine the {\em full} geometry of the spacetime in {\em finite} eigentime.  In other words:  can an observer determine the holonomies $h(\lambda)$ for a set of generators of  the fundamental group $\pi_1(M)$ up to conjugation in finite eigentime based on such measurements?
It is shown in  \cite{ichb,ichw} that this is indeed possible. By considering an observer who emits light in all directions at a given time and measures the return time and return direction for each returning  lightray, one obtains a procedure 
that yields  the holonomies  (up to conjugation)  from a {\em finite number} of returning lightrays
  \cite{ichb,ichw}. The measurements thus allow an observer to determine the associated point in  phase space $\mathcal P$ \eqref{phsp} and the full geometry of the spacetime  in finite eigentime.

{\large\bf Acknowledgements}

I thank the organisers of  {\em 2nd School and Workshop on
Quantum Gravity and Quantum Geometry}
(Corfu, 13-20/09/2009), where this work was presented as a talk. 
My research is supported by the Emmy Noether fellowship ME 3425/1-1 of the German Research Foundation (DFG).

\end{document}